\documentclass[preprint,prd,superscriptaddress]{revtex4}
\newcommand{\p}{\partial}
\newcommand{\pslash}{p\kern-1ex /}
\newcommand{\lslash}{l\kern-1ex /}
\newcommand{\kslash}{k\kern-1ex /}
\newcommand{\dslash}{\p\kern-1.2ex /}
\newcommand{\Dslash}{{\cal D}\kern-1.5ex /}
\newcommand{\Tr}{{\rm Tr}}
\newcommand{\tr}{{\rm tr}}
\newcommand{\re}{{\rm Re}}
\newcommand{\im}{{\rm Im}}
\def\Id{\mbox{1\hspace{-1.2mm}I} }

\newcommand{\diag}{{\rm diag}}
\newcommand{\sumover}[1]{\sum_{#1=1}^{N_f}}
\newcommand{\bea}{\begin{eqnarray}}
\newcommand{\eea}{\end{eqnarray}}
\newcommand{\vol}{\Omega}
\newcommand{\nn}{\nonumber\\}
\newcommand{\EQ}{\hspace{-2mm} &=& \hspace{-2mm}}
\newcommand{\EM}{\hspace{-2mm} & & \hspace{-2mm}}

\newcommand{\BAN}{\begin{eqnarray*}}
\newcommand{\EAN}{\end{eqnarray*}}
\begin{document}

\newcommand{\NTU}{
  Department of Physics, and Center for Theoretical Sciences, \\ 
  National Taiwan University, Taipei~10617, Taiwan  
}

\newcommand{\CQSE}{
  Center for Quantum Science and Engineering,  
  National Taiwan University, Taipei~10617, Taiwan  
}

\preprint{NTUTH-09-505A}
 
\title{Topological Susceptibility to the One-Loop Order in  \\ 
       Chiral Perturbation Theory}  

\author{Yao-Yuan Mao}
\affiliation{\NTU}

\author{Ting-Wai~Chiu}
\affiliation{\NTU}
\affiliation{\CQSE}

\collaboration{for the TWQCD Collaboration}
\noaffiliation


\begin{abstract}

We derive the topological susceptibility to the one-loop order 
in the chiral effective theory of QCD, for an arbitrary number of flavors. 

\end{abstract}

\maketitle

\section{Introduction}

In Quantum Chromodynamics (QCD), the topological susceptibility
($ \chi_t $) is the most crucial quantity to measure the
topological charge fluctuation of the QCD vacuum,
which plays an important role in breaking the $ U_A(1) $ symmetry.
Theoretically, $ \chi_t $ is defined as
\bea
\label{eq:chi_t}
\chi_{t} = \int d^4 x  \left< \rho(x) \rho(0) \right>, 
\eea
where 
\bea
\label{eq:rho}
\rho(x) = \frac{1}{32 \pi^2} \epsilon_{\mu\nu\lambda\sigma}
                             \tr[ F_{\mu\nu}(x) F_{\lambda\sigma}(x) ], 
\eea
is the topological charge density 
expressed in term of the matrix-valued field tensor $ F_{\mu\nu} $.
With mild assumptions, Witten \cite{Witten:1979vv} and
Veneziano \cite{Veneziano:1979ec}
obtained a relationship between the topological susceptibility
in the quenched approximation and the mass of $ \eta' $ 
meson (flavor singlet) in unquenched QCD with $ N_f $ degenerate flavors, 
namely, 
\BAN
\chi_t(\mbox{quenched}) = \frac{F_\pi^2 m_{\eta'}^2}{2 N_f}, 
\EAN
where $ F_\pi \simeq 93 $ MeV, the decay constant of pion.
This implies that the mass of $ \eta' $ is essentially due to
the axial anomaly relating to non-trivial topological charge
fluctuations, which can turn out to be nonzero even in the chiral limit,
unlike those of the (non-singlet) approximate Goldstone bosons.

Using the Chiral Perturbation Theory (ChPT), 
Leutwyler and Smilga \cite{Leutwyler:1992yt,Smilga:2001ck}
obtained the following relations in the chiral limit  
\bea
\label{eq:LS_nf2}
\chi_t \EQ \Sigma \left( \frac{1}{m_u} +\frac{1}{m_d} \right)^{-1}, 
\hspace{4mm} (N_f = 2),   \\ 
\label{eq:LS_nf2p1}
\chi_t \EQ \Sigma
         \left( \frac{1}{m_u} +\frac{1}{m_d} +\frac{1}{m_s} \right)^{-1}, 
\hspace{4mm} (N_f = 3), 
\eea
where $ m_u$, $m_d$, and $m_s$ are the quark masses, 
and $ \Sigma $ is the chiral condensate.
This implies that in the chiral limit ($ m_u \to 0 $) 
the topological susceptibility is suppressed due to internal quark loops.
Most importantly, (\ref{eq:LS_nf2}) and (\ref{eq:LS_nf2p1}) 
provide a viable way to extract $ \Sigma $ from $ \chi_t $ in the chiral limit.

From (\ref{eq:chi_t}), one obtains
\bea
\label{eq:chit_Qt}
\chi_t = \frac{\left< Q_t^2 \right>}{\vol}, \hspace{4mm}
Q_t \equiv  \int d^4 x \rho(x), 
\eea
where $ \vol $ is the volume of the system, and
$ Q_t $ is the topological charge (which is an integer for QCD).
Thus, one can determine $ \chi_t $ by counting the number of
gauge configurations for each topological sector.  
Furthermore, we can also obtain the second normalized cumulant 
\bea 
\label{eq:c4}
c_4 = -\frac{1}{\vol} \left[   \langle Q_t^4 \rangle
                            -3 \langle Q_t^2 \rangle^2 \right], 
\eea
which is related to the leading anomalous contribution to 
the $ \eta'-\eta' $ scattering amplitude in QCD, as well as the 
dependence of the vacuum energy on the vacuum angle $ \theta $.     
(For a recent review, see for example, Ref. \cite{Vicari:2008jw} 
and references therein.)

However, for lattice QCD, it is difficult to extract $ \rho(x) $
and $ Q_t $ unambiguously from the gauge link variables, due to
their rather strong fluctuations.

To circumvent this difficulty, one may consider
the Atiyah-Singer index theorem
\cite{Atiyah:1968mp}
\bea
\label{eq:AS_thm}
Q_t = n_+ - n_- = \mbox{index}({\cal D}),
\eea
where $ n_\pm $ is the number of zero modes of the massless Dirac
operator $ {\cal D} \equiv \gamma_\mu ( \partial_\mu + i g A_\mu) $
with $ \pm $ chirality. Since $ {\cal D} $ is anti-Hermitian and chirally
symmetric, its nonzero eigenmodes must come in complex conjugate pairs
with zero chirality. 
Thus one can obtain the identity
\bea
\label{eq:index_Q1}
n_+ - n_- = m \int d^4 x \ \tr [ \gamma_5 ({\cal D} + m)^{-1}(x,x)],
\eea
by spectral decomposition, where the nonzero modes drop out
due to zero chirality. In view of (\ref{eq:AS_thm}) and (\ref{eq:index_Q1}),
one can regard 
$ \rho_t(x) \equiv m_q \tr [ \gamma_5 ({\cal D} + m_q)^{-1}(x,x)] $
as topological charge density, to replace $ \rho(x) $
in the measurement of $ \chi_t $.

Recently, the topological susceptibility and the second normalized cumulant 
have been measured in unquenched lattice QCD with exact chiral symmetry, 
for $ N_f = 2 $  and $ N_f = 2+1 $ lattice QCD with overlap fermion 
in a fixed topology \cite{Aoki:2007pw, Chiu:2008kt},  
and $ N_f = 2+1 $ lattice QCD with domain-wall fermion 
\cite{Chiu:2008jq}. 
The results of topological susceptibility turn out in good agreement 
with the Leutwyler-Smilga relation, with the values of the chiral
condensate as follows.
\BAN   
\Sigma^{\overline{\mathrm{MS}}}(\mathrm{2~GeV})
&=& [\mathrm{245(5)(12)~MeV}]^3, \hspace{4mm} (N_f = 2),  \hspace{4mm} 
\mbox{Ref. [7]},  \\  
\Sigma^{\overline{\mathrm{MS}}}(\mathrm{2~GeV})
&=& [\mathrm{253(4)(6)~MeV}]^3, \hspace{4mm} (N_f = 2+1),  \hspace{4mm} 
\mbox{Ref. [8]},  \\  
\Sigma^{\overline{\mathrm{MS}}}(\mathrm{2~GeV})
&=& [\mathrm{259(6)(9)~MeV}]^3, \hspace{4mm} (N_f = 2+1),  \hspace{4mm} 
\mbox{Ref. [9]}. 
\EAN 
These results assure that lattice QCD with exact chiral symmetry is 
the proper framework to tackle the strong interaction physics
with topologically non-trivial vacuum fluctuations. 
Obviously, the next task for unquenched lattice QCD with exact chiral symmetry 
is to determine the second normalized cumulant $ c_4 $ to a good 
precision, and to address the question how the vacuum energy depends 
on the vacuum angle $ \theta $ and related problems. 
Theoretically, it is interesting to obtain an  
analytic expression of $ c_4 $ in ChPT, 
as well as to extend the Leutwyler-Smilga relation to the 
one-loop order. In this paper, we derive  
the topological susceptibility to the one-loop order 
in ChPT, for an arbitrary number of flavors. 

The outline of this paper is as follows. 
In Section 2, we review the derivation of topological susceptibility 
$ \chi_t $ at the tree level of ChPT, and also derive 
the second normalized cumulant $ c_4 $ 
at the tree level, and discuss its implications.   
In Section 3, we derive $ \chi_t $ up to the one-loop order in ChPT 
for an arbitrary number of flavors.    
In Section 4, we conclude with some remarks, and also present the case 
of $ 2+1 $ flavors, in which only the one-loop corrections
due to the $ u $ and $ d $ quarks are incorporated. 
In the Appendix, we present a heuristic derivation of  
the counterpart of the Leutwyler-Smilga relation 
in lattice QCD with exact chiral symmetry.

\section{Topological susceptibility at the tree level of ChPT}

Before we proceed to derive $ \chi_t $ to the one-loop order in ChPT, 
it is instructive for us to recap the derivation of $ \chi_t $ 
at the tree level \cite{Leutwyler:1992yt,Smilga:2001ck}. 

The leading terms of the effective chiral lagrangian for QCD 
with $ N_f $ flavor at $ \theta = 0 $ \cite{Gasser:1984gg}
are the kinetic term and the symmetry breaking term,  
\bea
\label{eq:L2}
\mathcal{L}^{(2)} = \mathcal{L}_{\text{eff}}^{(2)} + 
                    \mathcal{L}_{\text{s.b.}}^{(2)}
= \frac{F_\pi^2}{4} \Tr(\partial_\mu U \partial^\mu U^\dagger) 
+ \frac{\Sigma}{2} \Tr(\mathcal{M} U^\dagger + U \mathcal{M}^\dagger), 
\eea
where $ U(x) = \exp\{ 2 i \phi^a(x) t^a/F_\pi \} $ is a group element of 
$ SU(N_f) $, $ \mathcal{M} $ is the quark mass matrix,      
$ F_\pi $ is the pion decay constant, and 
$ \Sigma = \langle \bar\psi \psi \rangle_{\text vac} $ 
is the chiral condensate of the QCD vacuum.

On the other hand, the partition function of QCD in the $\theta$ vacuum 
can be written as 
\bea
\label{eq:Z_Nf}
Z_{N_f}(\theta) = \sum_{Q} e^{-iQ\theta} Z_Q,
\eea
where
\BAN
Z_Q &=& \int [d A_\mu] e^{-S_G[A_\mu]} \det \left( \gamma_\mu D_\mu  
        + \frac{1-\gamma^5}{2} \mathcal{M} 
        + \frac{1+\gamma^5}{2} \mathcal{M}^\dagger \right)  \\
&=& \int [d A_\mu ] e^{-S_G[A_\mu]} 
    \prod_{k} \det(\lambda_k^2 + \mathcal{M}^\dagger \mathcal{M}) \cdot
\left\{  
\begin{array}{c}
\det(\mathcal{M}^\dagger)^{Q}, \hspace{4mm} Q > 0,  \\ 
\det(\mathcal{M})^{-Q}, \hspace{4mm} Q < 0,   
\end{array} \right.
\EAN
where $ S_G $ is the action of the gauge field, 
and $ \lambda_k $'s are non-zero eigenvalues of the 
massless Dirac operator $ \gamma_\mu D_\mu $ in the gauge background. 
Thus the physical vacuum angle on which all physical quantities depend is  
$ \theta_{\text{phys}} = \theta + \mbox{arg} \det(\mathcal{M}) $
rather than $ \theta $. Also, the $\theta$-dependence of 
$ Z_{N_f}(\theta) $ always enters through the combinations 
$ {\cal M} e^{i \theta /N_f} $ and $ {\cal M}^\dagger e^{-i \theta /N_f} $.  
It follows that for $ \theta \ne 0 $, the symmetry breaking term 
in the chiral effective lagrangian can be written as
\bea
\label{eq:L_sb}
  \mathcal{L}_{\text{s.b.}}^{(2)} = \Sigma \ 
\mbox{Re} \left[ \Tr(\mathcal{M} e^{i \theta / N_f } U^\dagger) \right].
\eea

Defining the vacuum energy density
\bea
\label{eq:vacuum_energy}
\epsilon_{\text{vac}}(\mathcal{M}, \theta) = 
-\frac{1}{\vol} \log Z_{N_f}(\theta), 
\eea
then the topological susceptibility $ \chi_t $ (\ref{eq:chit_Qt})
and the second normalized culmulant $ c_4 $ (\ref{eq:c4}) 
can be expressed as  
\bea
\label{eq:chit_vac}
\chi_t &=& \left. 
           \frac{ \partial^2 \epsilon_{\text{vac}}(\mathcal{M}, \theta)}
              { \partial^2 \theta} \right|_{\theta = 0},     \\
\label{eq:c4_vac}
c_4 &=& \left.\frac{ \partial^4 \epsilon_{\text{vac}}(\mathcal{M}, \theta)}
              { \partial^4 \theta} \right|_{\theta = 0}.    
\eea 

For small quark masses ($ L \ll m_\pi^{-1} $), the unitary matrix $ U $ 
does not depend on $ x_\mu $. Thus only the symmetry-breaking term survives 
in (\ref{eq:L2}), and the partition function becomes 
\bea
\label{eq:partition}
Z_{N_f}(\theta) = \int dU \exp \left\{ 
\vol \ \Sigma \ {\text Re} 
\left[ \Tr(\mathcal{M} e^{i\theta/N_f} U^\dagger) \right] \right\}, 
\eea
where $ \vol = L^3 T $ is the space-time volume. 
Without loss of generality, the unitary matrix $ U $ can be taken 
to be diagonal
\BAN
U = \diag \left(e^{i\alpha_1}, e^{i\alpha_2}, \dots, e^{i\alpha_{N_f}} \right), \hspace{4mm} 
\sumover{j} \alpha_{j} = 0, 
\EAN
where the last constraint follows from the unitarity of $ U $.  
Also, we can choose the mass matrix to be diagonal 
$ \mathcal{M} = \diag(m_1, \dots, m_{N_f}) $. 
Then we have 
\BAN
\re \left[ \Tr(\mathcal{M} e^{i\theta/N_f} U^\dagger) \right] 
= \sum_j m_j \cos \phi_j, 
\EAN
where $ \phi_j = \theta/N_f - \alpha_j $, and  
$ \sum_j \phi_j = \theta $.

Now, we consider a sufficiently large volume $ \vol $ satisfying 
$ m_j \Sigma \vol \gg 1 $, 
then the group integral in the partition function (\ref{eq:partition})
is largely due to the $ U $ which minimizes the minus exponent of the 
integrand, i.e.,   
\bea
\label{eq:min}
\min_{U} \left\lbrace - \re 
\left[ \Tr(\mathcal{M} e^{i\theta/N_f} U^\dagger )\right] \right\rbrace
= \min_{\phi} \left\{ - \sumover{j}{m_j \cos \phi_j} \right\}, 
\hspace{4mm}
\sumover{j} \phi_j = \theta.
\eea

For $ N_f = 2 $, this amounts to minimize the function 
\BAN
\label{eq:min_nf2}
- m_1 \cos(\phi_1) - m_2 \cos(\theta - \phi_1),  
\EAN
where the constraint $ \phi_1 + \phi_2 = \theta $ has been used. 
A simple calculation gives the minimum,   
\BAN
- \sqrt{ m_1^2 + m_2^2 + 2 m_1 m_2 \cos\theta}. 
\EAN  
Thus the partition function is 
\BAN
Z_{N_f}(\theta) = Z_0 \exp \left\{ 
\vol \Sigma   
\sqrt{ m_1^2 + m_2^2 + 2 m_1 m_2 \cos\theta} 
\right\}, \hspace{4mm}  m \Sigma \vol \gg 1, 
\EAN
which gives the vacuum energy
\BAN
\label{eq:e_vac}
\epsilon_{\text{vac}}(\theta) = \epsilon_0 
-\Sigma \sqrt{m_u^2+m_d^2+2m_u m_d \cos \theta}, 
\EAN
where $ \epsilon_0 $ is the additive normalization constant 
corresponding the normalization factor $ Z_0 $ in the partition function. 
From (\ref{eq:e_vac}), we obtain the topological susceptibility 
\bea
\label{eq:chit_nf2_tree}
\chi_t = \frac{\partial^2 \epsilon_{\text{vac}}}{\partial \theta^2} 
         \bigg \vert_{\theta = 0} = \Sigma \frac{m_u m_d}{m_u + m_d}. 
\eea
Furthermore, the second normalized cumulant is
\bea
\label{eq:c4_nf2_tree}
c_4 = \frac{\partial^4 \epsilon_{\text{vac}}}{\partial \theta^4} 
        \bigg \vert_{\theta = 0} 
= - \Sigma \frac{m_u m_d}{m_u+m_d} + 3\Sigma \frac{m_u^2 m_d^2}{(m_u+m_d)^3}
= - \Sigma \left( \frac{1}{m_u^3} + \frac{1}{m_d^3} \right) 
    \left( \frac{m_u m_d}{m_u+m_d} \right)^4 , 
\eea
which has not been discussed explicitly in the literature. 
The vital observation is that the ratio of $ \chi_t $ and $ c_4 $ is 
\bea
\label{eq:c4_chit_nf2_tree}
\frac{c_4}{\chi_t} = - 1 + \frac{3m_um_d}{(m_u+m_d)^2},    
\eea
which goes to $ -1/4 $ in the isospin limit $ m_u = m_d $. 
This seems to rule out the dilute instanton gas/liquid model
\cite{'tHooft:1976fv, Callan:1977gz, Shuryak:1981ff}
which predicts that $ c_4 / \chi_t = -1 $. 
Moreover, recent numerical results of $ c_4/\chi_t $ from 
quenched lattice QCD \cite{Giusti:2007tu,DelDebbio:2007kz,Durr:2006ky}
and unquenched lattice QCD \cite{Chiu:2008kt,Chiu:2008jq}
are consistent with the prediction of ChPT.   

Next we turn to the case $ N_f > 2 $. 
Then there is no analytic solution to the minimization problem (\ref{eq:min}).
However, for the purpose of obtaining the topological susceptibility, 
one may consider the limit of small $ \theta $ (and $ \phi_j $'s)  
because $ U = \Id $ gives the minimal vacuum energy at $ \theta = 0 $. 
Since $ \chi_t $ only depends on the curvature of 
$ \epsilon_{\text{vac}}(\theta) $ around $ \theta = 0 $, 
this approximation would give the exact result of $ \chi_t $ (at the tree-level).
To the order of $ \theta^4 $, the minimization problem (\ref{eq:min}) becomes 
\BAN
\label{eq:V_eff}
\min_{\phi} \left\{ - \sumover{j}{m_j \cos \phi_j} \right\}
= \min_{\phi} \left\{ \frac{1}{2} \sumover{j} m_j \phi_j^2 
	- \frac{1}{24} \sumover{j} m_j \phi_j^4 \right\}, 
\hspace{4mm}
\sumover{i} \phi_i = \theta.
\EAN
Now introducing the Lagrange multiplier $ \lambda $ 
to incorporate the constraint $ \sum_i \phi_i = \theta $, 
then the minimization problem amounts to solving the equation    
\BAN
\label{eq:min_condition_1}
\frac{\partial}{\partial \phi_i} 
\left[ \frac{1}{2} \sumover{j} m_j \phi_j^2 
	- \frac{1}{24} \sumover{j} m_j \phi_j^4 
    - \lambda \left( \sumover{j} \phi_j - \theta \right) \right] 
= m_i \phi_i - \frac{1}{6} m_i \phi_i^3 - \lambda = 0.
\EAN
Setting $ \phi_i = a_1 \frac{\lambda}{m_i} + a_3 \left(\frac{\lambda}{m_i} \right)^3$ 
(where $ a_1 $ and $ a_3 $ are parameters), and using $ \sum_i \phi_i = \theta $, 
we can solve for $ a_1 $ and $ a_3 $, and $ \phi_i $ to the order of $ \theta^3 $, 
\BAN
\phi_i = \frac{\bar{m}}{m_i} \theta 
+ \frac{\theta^3}{6}
  \left[ \left( \frac{\bar{m}}{m_i} \right)^3 - 
 	     \left( \frac{\bar{m}}{m_i} \right)
 	     \sumover{j} \left( \frac{\bar{m}}{m_j} \right)^3 \right]
+ \mathcal{O}(\theta^5).
\EAN
where $ {\bar m} \equiv \left(\sumover{i} m_i^{-1} \right)^{-1} $ is the 
``reduced mass" of the $ N_f $ quark flavors. 
Keeping the exponent of the partition function to the order of $ \theta^4 $, we have 
\bea
Z_{N_f}(\theta) = Z_0 \exp \left\{ - \vol \Sigma 
               \left( \sum_{j=1}^{N_f} \frac{1}{m_j} \right)^{-1} 
               \frac{\theta^2}{2} + 
               \vol \Sigma 
               \sum_{i=1}^{N_f} m_i^{-3}
               \left( \sum_{j=1}^{N_f} \frac{1}{m_j} \right)^{-4} 
               \frac{\theta^4}{24} + 
               {\cal O}(\theta^6) \right\},
\eea
and the vacuum energy density is  
\BAN
\epsilon_{\text{vac}}(\theta) = \epsilon_0 + 
  \Sigma \left(\sum_{j=1}^{N_f} \frac{1}{m_j} \right)^{-1} \frac{\theta^2}{2}       
  - \Sigma \sum_{i=1}^{N_f} m_i^{-3}
           \left( \sum_{j=1}^{N_f} \frac{1}{m_j} \right)^{-4} 
           \frac{\theta^4}{24} 
  + {\cal O}(\theta^6) . 
\EAN
It follows that the topological susceptibility is 
\bea
\label{eq:chit_nf_tree}
\chi_t = \frac{\partial^2 \epsilon_{\text{vac}}}{\partial \theta^2} 
         \bigg \vert_{\theta = 0} 
       = \Sigma \left( \sum_{j=1}^{N_f} \frac{1}{m_j} \right)^{-1}, 
\eea
and 
\bea
\label{eq:c4_nf_tree}
c_4 = \frac{\partial^4 \epsilon_{\text{vac}}}{\partial \theta^4} 
         \bigg \vert_{\theta = 0} 
    = - \Sigma \sum_{i=1}^{N_f} m_i^{-3}
        \left( \sum_{j=1}^{N_f} \frac{1}{m_j} \right)^{-4} , 
\eea
which generalize Eqs.~(\ref{eq:chit_nf2_tree}) and (\ref{eq:c4_nf2_tree}) to 
an arbitrary number of flavors. In particular, for $ N_f = 3 $, 
\bea
\label{eq:chit_tree}
\chi_t=\Sigma\left(\frac{1}{m_u}+\frac{1}{m_d}+\frac{1}{m_s}\right)^{-1},
\eea
and  
\bea
\label{eq:c4_nf3_tree}
c_4 = - \Sigma 
\left( \frac{1}{m_u^3} + \frac{1}{m_d^3} + \frac{1}{m_s^3} \right) 
\left( \frac{1}{m_u} + \frac{1}{m_d} + \frac{1}{m_s} \right)^{-4} .
\eea
Nevertheless, these two formulas seem unnatural, since the strange quark
is much heavier than the up and down quarks. Thus a plausible 
chiral limit is to take $ m_{u,d} \to 0 $, while keeping $ m_s $ fixed. 
Consequently, the condensate of the strange quark 
$ \langle \bar s s \rangle $ must be different from $ \Sigma $, 
and it should also enter this formula.   
In the Appendix, we present a heuristic derivation 
of the counterpart of (\ref{eq:chit_tree})    
in lattice QCD with exact chiral symmetry, which 
takes into account of the difference between 
$ \langle \bar s s \rangle $ and $ \Sigma $, 
as given in Eq. (\ref{eq:chit_nf2p1_qcd}).

\section{Topological susceptibility to the one-loop order of ChPT}

To the one-loop order of ChPT, one has to include 
$ \mathcal{L}^{(4)} $ \cite{Gasser:1984gg}
at the tree level as well as the one-loop contributions 
of $ \mathcal{L}^{(2)} $. 
In 1984, Gasser and Leutwyler \cite{Gasser:1984gg} 
considered the low-energy expansion,
where both $ p $ and $ \cal M $ are assumed to be small
but $ {\cal M}/p^2 $ can have a finite value,
such that the value of $ M^2_\pi/p^2 $ can be fixed.
In this case, the external sources
$ a_\mu(x) $ and $ p(x) $ can be counted as order of $ \Phi $, 
and $ v_\mu(x) $ and $ s(x) - {\cal M} $ as order of $ \Phi^2 $.
Gasser and Leutwyler showed that at the one-loop order, 
the chiral effective action can be written as
\bea
W = W_t + W_u + W_A + \mathcal{O}(\Phi^6) ,
\eea
where $W_t$ denotes the sum of tree diagrams and tadpole contributions 
(of order $ \Phi^2 $), 
$ W_u $ the unitarity correction (of order $ \Phi^3 $), 
and $ W_A $ the anomaly contribution (of order $ \Phi^4 $).
Because the $ \theta $ dependence enters the
Lagrangian only through $ \cal M $, 
we can count $ \chi_t $ as order of $ \Phi^2 $,
thus for the evaluation of topological susceptibility to the one-loop order, 
and it suffices to consider $ W_t $ only.

Moreover, Gasser and Leutwyler \cite{Gasser:1984gg} showed that  
the pole terms due to the
one-loop contributions of $ \mathcal{L}^{(2)} $ can be absorbed by 
the low-energy coupling constants of $ \mathcal{L}^{(4)} $, 
and $W_t$ is given by \cite{Gasser:1984gg}
\bea
\label{eq:Wt}
W_t \EQ \sum_P \int d^4 x \, \frac{F_\pi^2}{2} 
\left\lbrace \frac{1}{N_f} - \frac{M_P^2}{16\pi^2 F_\pi^2} 
\ln \frac{M_P^2}{\mu_{sub}^2}
\right\rbrace \sigma_{PP}^{\Delta} \nn*
\EM + \sum_P \int d^4 x \, \frac{F_\pi^2}{2} 
\left\lbrace \frac{N_f}{N_f^2-1} - 
\frac{M_P^2}{16 \pi^2 F_\pi^2} \ln \frac{M_P^2}{\mu_{sub}^2}  
\right\rbrace \sigma_{PP}^{\chi}
+ \int d^4 x \mathcal{L}^{r(4)} , 
\eea
where $ M_P^2 $'s are the squared meson masses, $ \sigma_{PP}^{\Delta} $
corresponds to the kinetic term which can be dropped in the limit of 
small quark masses, $ \sigma_{PP}^{\chi} $ corresponds to 
the symmetry breaking term,  
\bea
\sigma_{PP}^{\chi} = \frac{1}{8} 
\Tr\left( \left\lbrace \lambda_P , \lambda_P^\dagger \right\rbrace 
(\chi^\dagger U + U^\dagger \chi) \right) - M_P^2 \ , 
\eea
and $ \mathcal{L}^{r(4)} $ is just
$ \mathcal{L}^{(4)} $ with renormalized low-energy coupling constants,  
\bea
\mathcal{L}^{r(4)} 
\EQ L_1^r \left\{\mbox{Tr}[D_{\mu}U (D^{\mu}U)^{\dagger}] \right\}^2
+ L_2^r \Tr \left [D_{\mu}U (D_{\nu}U)^{\dagger}\right]
\Tr \left [D^{\mu}U (D^{\nu}U)^{\dagger}\right] \nn*
\EM + L_3^r \Tr\left[ 
D_{\mu}U (D^{\mu}U)^{\dagger}D_{\nu}U (D^{\nu}U)^{\dagger}
\right ] \nn*
\EM + L_4^r \Tr \left [ D_{\mu}U (D^{\mu}U)^{\dagger} \right ]
\Tr \left( \chi U^{\dagger}+ U \chi^{\dagger} \right ) \nn*
\EM + L_5^r \Tr \left[ D_{\mu}U (D^{\mu}U)^{\dagger}
(\chi U^{\dagger}+ U \chi^{\dagger})\right]
+ L_6^r \left[ \Tr \left ( \chi U^{\dagger}+ U \chi^{\dagger} \right )
\right]^2 \nn*
\EM + L_7^r \left[ \Tr \left ( \chi U^{\dagger} - U \chi^{\dagger} \right )
\right]^2
+ L_8^r \Tr \left ( U \chi^{\dagger} U \chi^{\dagger}
+ \chi U^{\dagger} \chi U^{\dagger} \right ) \nn*
\EM - i L_9^r \Tr \left [ F^R_{\mu\nu} D^{\mu} U (D^{\nu} U)^{\dagger}
+ F^L_{\mu\nu} (D^{\mu} U)^{\dagger} D^{\nu} U \right ]
+ L_{10}^r \Tr \left ( U F^L_{\mu\nu} U^{\dagger} F_R^{\mu\nu} \right) \nn*
\EM + H_1^r \Tr \left ( F^R_{\mu\nu} F^{\mu\nu}_R +
F^L_{\mu\nu} F^{\mu\nu}_L \right )
+ H_2^r \Tr \left( \chi \chi^{\dagger} \right) .
\eea
Here $ \chi = 2 (\Sigma/F_\pi^2) \mathcal{M} \equiv 2 B_0 {\cal M} $, 
$ \lambda_P $'s are the generators of $ SU(N) $ in the physical basis,
$ \{ L_i^r(\mu_{sub}), i=1,\cdots, 10 \} $ are 
renormalized low-energy coupling constants, 
and the last two contact terms (with couplings $ H_1^r(\mu_{sub}) $ 
and $ H_2^r(\mu_{sub}) $) are the counter terms required for 
renormalization of the one-loop diagrams.   

For small quark masses ($ L \ll m_\pi^{-1} $), the unitary matrix $ U $ 
does not depend on $ x_\mu $, thus the term involving 
$ \sigma_{PP}^{\Delta} $ in (\ref{eq:Wt}) can be dropped. 

Next we consider the term with $ \sigma_{PP}^{\chi} $ in (\ref{eq:Wt}).
Using the formula 
\bea
\label{eq:Nf}  
\sum_P \left\lbrace \lambda_P , \lambda_P^\dagger \right\rbrace 
= \frac{4(N_f^2-1)}{N_f} \, \Id ,  
\eea 
we obtain its contribution to the chiral effective lagrangian,  
\bea
\label{eq:L2_tree_tadpole}
&&  \Sigma \re \Tr (\mathcal{M}U^\dagger) 
   - \frac{N_f F_\pi^2}{2(N_f^2-1)} \sum_P M_P^2    \nn
&& - \frac{\Sigma}{4 F_\pi^2} \sum_P 
    \re \Tr \left( \left\lbrace \lambda_P, \lambda_P^\dagger \right\rbrace
 	{\cal M} U^\dagger \right) 
 	\frac{M_P^2}{16 \pi^2} \ln \frac{M_P^2}{\mu_{sub}^2}
+ \frac{M_P^4}{32 \pi^2} \ln \frac{M_P^2}{\mu_{sub}^2}.
\eea
For small quark masses ($ L \ll m_\pi^{-1} $), the unitary matrix $ U $
does not depend on $ x_\mu $, so only the sixth, seventh, and eighth terms
in $ \mathcal{L}^{r(4)} $ are relevant to the partition function.
For $ \theta \ne 0 $, $ \theta $ enters the chiral effective lagrangian
through the combinations
$ \mathcal{M} e^{i\theta/N_f} $ and $ \mathcal{M}^\dagger e^{-i\theta/N_f} $.
Thus these three potential terms can be written as
\bea
\label{eq:L_678}
L_6^r \left[ 4 B_0 \re\Tr(\mathcal{M} e^{i\theta/N_f} U^\dagger) \right]^2
+ L_7^r \left[ i 4 B_0 \im\Tr(\mathcal{M} e^{i\theta/N_f} U^\dagger) \right]^2
+8 L_8^r B_0^2 \re\Tr\left[(\mathcal{M} e^{i\theta/N_f} U^\dagger)^2 \right]. 
\nn
\eea

Without loss of generality, we can take $ U $ and $ {\cal M} $
to be diagonal,  
\BAN
U &=& \diag \left(e^{i\alpha_1}, e^{i\alpha_2}, 
\dots, e^{i\alpha_{N_f}} \right), \hspace{4mm} 
\sumover{j} \alpha_{j} = 0, \\ 
\mathcal{M} &=& \diag(m_1, \dots, m_{N_f}),  
\EAN
therefore the contributions of (\ref{eq:L2_tree_tadpole})  
and (\ref{eq:L_678}) become (dropping the terms without $ U $ dependence)
\BAN
&\quad&\Sigma \, \sumover{j} m_j \cos \phi_j 
 -  \frac{\Sigma}{4 F_\pi^2} \sum_P \sumover{j} 
	\left\lbrace \lambda_P, \lambda_P^\dagger \right\rbrace_{jj} 
	m_j \cos \phi_j
	\frac{M_P^2}{16 \pi^2} \ln \frac{M_P^2}{\mu_{sub}^2} \nn*
&+& 16 B_0^2 L_6^r \left( \sumover{j} m_j \cos \phi_j \right)^2 
 - 16 B_0^2 L_7^r \left( \sumover{j} m_j \sin \phi_j \right)^2 
 + 8 B_0^2 L_8^r \sumover{j} m_j^2 \cos 2\phi_j, 
\EAN
where $ \phi_j = \theta/N_f - \alpha_j $, and  $ \sum_j \phi_j = \theta $.

Again, we use small $ \theta $ (small $ \phi_j $'s) approximation 
and keep terms up to the order of $ \phi_j^2 $, then the 
evaluation of the integral in the partition function in the limit 
$ m_j \vol \Sigma \gg 1 $ amounts to minimizing the generating functional    
\bea
\label{eq:1loop_min}
\EM \min_{\phi} \Bigg[ 
\frac{\Sigma}{2} \sumover{j} m_j \phi_j^2
-   \frac{\Sigma}{8 F_\pi^2} \sum_P \sumover{j} 
	\left\lbrace \lambda_P, \lambda_P^\dagger \right\rbrace_{jj} m_j \phi_j^2
    \frac{M_P^2}{16 \pi^2} \ln \frac{M_P^2}{\mu_{sub}^2} \nn*
\EM \quad + 16 B_0^2 L_6^r  \sumover{i} m_i \sumover{j} m_j \phi_j^2    
+ 16 B_0^2 L_7^r \left( \sumover{j} m_j \phi_j \right)^2 
+ 16 B_0^2 L_8^r \sumover{j} m_j^2 \phi_j^2 \Bigg],
\eea
with the constraint $ \sum_j \phi_j = \theta $.
We introduce the Lagrange multiplier $ \lambda $ to incorporate the
constraint in finding the minimum. For simplicity, we define
\BAN
A_j & \equiv & \frac{\Sigma}{2} m_j 
-   \frac{\Sigma}{8 F_\pi^2}
    \sum_P \left\lbrace \lambda_P, \lambda_P^\dagger \right\rbrace_{jj} m_j
	\frac{M_P^2}{16 \pi^2} \ln \frac{M_P^2}{\mu_{sub}^2} 
+   16 B_0^2 \left( L_6^r m_j \sumover{i} m_i + L_8^r m_j^2 \right), \\
B_j & \equiv & 4B_0 (L_7^r)^{1/2} m_j. 
\EAN
Then the minimization problem amounts to solving the equation 
\BAN
\frac{\partial}{\partial \phi_i} \left[ 
\sumover{j} A_j \phi_j^2 
+ \left( \sumover{j} B_j \phi_j \right)^2 
- \lambda \left( \sumover{j} \phi_j - \theta \right) \right] = 0, 
\EAN
which gives
\bea
\label{eq:min_condition}
A_i \phi_i + B_i \left( \sumover{j} B_j \phi_j \right) = \frac{\lambda}{2}.
\eea
Defining $ (\mathbf{T})_{ij} \equiv 2 A_i \delta_{ij} + 2 B_i B_j $, 
(\ref{eq:min_condition}) becomes
\bea
\sumover{j} (\mathbf{T})_{ij} \phi_j = \lambda, \quad i = 1, \dots, N_f.
\eea
Thus we can obtain $ \lambda $ using the constraint
\bea
\label{eq:lambda}
\theta = \sumover{j} \phi_j = 
\lambda \sumover{j} \sumover{i} (\mathbf{T}^{-1})_{ij} \ \Rightarrow \
\lambda =  \theta  
\left[ \sumover{j} \sumover{i} (\mathbf{T}^{-1})_{ij} \right]^{-1}.
\eea
Now multiplying Eq.~(\ref{eq:min_condition}) with $ \phi_i $ and summing 
over $ i $, we obtain 
\bea
\min_{\phi} \left[ \sumover{j} A_j \phi_j^2 
+ \left( \sumover{j} B_j \phi_j \right)^2 \right] = \frac{\lambda \theta}{2}, 
\eea
which can be used to simplify (\ref{eq:1loop_min}) to 
\bea
\frac{\lambda \theta}{2}  
=\frac{\theta^2}{2} \left[\sumover{i, j} (\mathbf{T}^{-1})_{ij} \right]^{-1}, 
\eea 
where the last equality follows from (\ref{eq:lambda}).
Finally, the partition function of QCD with $ N_f $ flavors
to the one-loop order of ChPT in the limit $ m \Sigma \vol \gg 1 $ is
equal to  
\bea
Z_{N_f}(\theta) = Z_0 \exp \left\{ \vol
\frac{\theta^2}{2} \left[\sumover{i, j} (\mathbf{T}^{-1})_{ij} \right]^{-1}
\right\}, 
\eea
and the vacuum energy density is 
\bea
\epsilon_{\text{vac}}(\theta)  
= \epsilon_0 
+ \frac{\theta^2}{2} \left[ \sumover{i, j} (\mathbf{T}^{-1})_{ij} \right]^{-1}. 
\eea
Thus the topological susceptibility to the one-loop order of ChPT is   
\bea
\chi_t = \frac{\partial^2 \epsilon_{\text{vac}}}{\partial \theta^2} 
         \bigg \vert_{\theta = 0} 
= \left[ \sumover{i, j} (\mathbf{T}^{-1})_{ij} \right]^{-1}.
\eea

To simplify the expression of topological susceptibility, 
we rewrite the matrix $ \mathbf{T} $ as 
\bea
(\mathbf{T})_{ij} \equiv 2 A_i \delta_{ij} + 2 B_i B_j = 
\Sigma (\mathcal{M} + \mathbf{T}')_{ij} ,
\eea
where 
\bea
\label{eq:matrix_T}
(\mathbf{T}')_{ij}  
\EQ  -\frac{1}{4 F_\pi^2} \sum_P 
\left\lbrace \lambda_P, \lambda_P^\dagger \right\rbrace_{jj}
 m_j \delta_{ij} 
\frac{M_P^2}{16 \pi^2} \ln \frac{M_P^2}{\mu_{sub}^2} \nn*
\EM  + K_6 \sumover{k} m_k m_j \delta_{ij}  
     +  K_8 m_j^2 \delta_{ij}  + K_7 m_i m_j,  
\eea
and
\bea
K_i \equiv \frac{32 B_0^2 L_i^r(\mu_{sub})}{\Sigma} 
= 32 \left(\frac{\Sigma}{F_\pi^4} \right) L_i^r(\mu_{sub}). 
\eea

Since the eigenvalues of the real and symmetric matrix
$ {\cal M}^{-1/2} \mathbf{T}' {\cal M}^{-1/2} $ are much less than 
one in the chiral limit, we can use the Taylor expansion   
$$ ( \Id + {\cal M}^{-1/2} \mathbf{T}' {\cal M}^{-1/2} )^{-1} \simeq    
    \Id - {\cal M}^{-1/2} \mathbf{T}' {\cal M}^{-1/2} + {\cal O}(m^2), $$    
and obtain
\bea
\label{eq:chit_nf_1loop}
\chi_t = \left[ \sumover{i, j} (\mathbf{T}^{-1})_{ij} \right]^{-1} 
& \simeq & \Sigma {\bar m} \left[ 1 + {\bar m} \sumover{i, j} \frac{(\mathbf{T}')_{ij}}{m_i m_j} \right] \nn*
& = & \Sigma \bar{m} \Bigg\lbrace 1 - \frac{1}{4 F_\pi^2}
\sum_P \sumover{j} 
\left\lbrace \lambda_P, \lambda_P^\dagger \right\rbrace_{jj}
\left( \frac{{\bar m}}{m_j} \right) 
\frac{M_P^2}{16 \pi^2} \ln \frac{M_P^2}{\mu_{sub}^2}  \nn* 
& \quad & \quad +  K_6 \sumover{i} m_i
+ N_f (N_f K_7 + K_8) {\bar m} \Bigg\rbrace , 
\eea
where 
$ {\bar m} \equiv \left(\sumover{i} m_i^{-1} \right)^{-1}$,
and all terms proportional to $ K_i^2 $ or $ K_i K_j $ have been dropped.
Equation (\ref{eq:chit_nf_1loop}) is the main result of this paper.  

For $ N_f = 2 $, there are three mesons, $ \pi^+ $, $\pi^0$, and $\pi^- $.
If we take their masses to be the same and use (\ref{eq:Nf}), we obtain
\bea
\label{eq:chit_nf2}
\chi_t
= \Sigma \left( \frac{1}{m_u} + \frac{1}{m_d} \right)^{-1} 
\bigg[1 &-& \frac{3}{2 F_\pi^2} 
\frac{M_\pi^2}{16 \pi^2} \ln \frac{M_\pi^2}{\mu_{sub}^2} 
+ K_6(m_u +m_d) \nn* 
&+&  2(2 K_7+K_8) \frac{m_u m_d}{m_u+m_d} \bigg] .
\eea

Next we turn to the case $ N_f = 3 $. Taking the eight 
pseudoscalar mesons with non-degenerate masses, we obtain
\bea
\label{eq:chit_nf3}
\chi_t 
& = & \Sigma {\bar m} \Bigg\lbrace 1 - \frac{1}{2F_\pi^2} \Bigg[ 
\sum_{i \neq j} 
\left( \frac{\bar{m}}{m_i} + \frac{\bar{m}}{m_j} \right) 
\frac{B_0(m_i+m_j)}{16 \pi^2} \ln \frac{B_0(m_i+m_j)}{\mu_{sub}^2} \nn*
& \quad & \qquad 
+ \left( \frac{\bar{m}}{m_u} + \frac{\bar{m}}{m_d} \right) 
\frac{M_{\pi^0}^2}{16 \pi^2} \ln \frac{M_{\pi^0}^2}{\mu_{sub}^2}
 + \frac{1}{3} 
\left( \frac{\bar{m}}{m_u} + \frac{\bar{m}}{m_d}+4\frac{{\bar m}}{m_s} \right) 
\frac{M_{\eta}^2}{16 \pi^2} \ln \frac{M_{\eta}^2}{\mu_{sub}^2}
\Bigg] \nn* 
& \quad & \qquad +  K_6 ( m_u + m_d + m_s )
+ 3 (3 K_7 + K_8) \bar{m} \Bigg\rbrace,
\eea
where $ \bar m = \left( m_u^{-1} + m_d^{-1} + m_s^{-1} \right)^{-1} $, 
and $ B_0 = \Sigma/F_\pi^2 $.

\section{Concluding Remark}

In this paper, we have derived the topological susceptibility 
to the one-loop order in ChPT, in the 
limit $ m \Sigma \vol \gg 1 $, for $ N_f = 2 $ [Eq. (\ref{eq:chit_nf2})],  
$ N_f = 3 $ [Eq. (\ref{eq:chit_nf3})], and an arbitrary number of 
flavors $ N_f $ [Eq. (\ref{eq:chit_nf_1loop})] respectively.   
     
For $ N_f = 3 $, since the mass of the strange quark is much heavier than 
the masses of $ u $ and $ d $ quarks, it seems reasonable just to 
incorporate the one-loop corrections due to the $ u $ and $ d $ quarks. 
Then, for $ N_f = 2+1 $ ($ u $ and $ d $ quarks to the one-loop order, 
and $ s $ quark at the tree level), the topological susceptibility becomes   
\bea
\label{eq:chit_nf2p1}
\chi_t
= \Sigma \Bigg\{
   \left(\frac{1}{m_u}+\frac{1}{m_d}\right) 
   \bigg[1 &+& \frac{3}{2 F_\pi^2}  \frac{M_\pi^2}{16 \pi^2} 
	\ln \frac{M_\pi^2}{\mu_{sub}^2} 
	- K_6(m_u +m_d) \nn* 
	&-&  2(2 K_7+K_8) \frac{m_u m_d}{m_u+m_d} \bigg] + 
   \frac{1}{m_s} \Bigg\}^{-1}. 
\eea
This supplements (\ref{eq:chit_nf3}) for the case $ N_f = 2+1 $. 

Now the trend of unquenched lattice QCD simulations is to  
include the charm quark. Thus it is also interesting 
to include the case $ N_f = 2+1+1 $, with both $ s $ and $ c $ quarks 
being kept at the tree level. Then the topological susceptibility is     
\bea
\label{eq:chit_nf2p1p1}
\chi_t = \Sigma \Bigg\{
   \left(\frac{1}{m_u}+\frac{1}{m_d}\right) 
   \bigg[1 &+& \frac{3}{2 F_\pi^2} \frac{M_\pi^2}{16 \pi^2} 
	\ln \frac{M_\pi^2}{\mu_{sub}^2} 
	- K_6(m_u +m_d) \nn* 
	&-&  2(2 K_7+K_8) \frac{m_u m_d}{m_u+m_d} \bigg] + 
   \frac{1}{m_s} + \frac{1}{m_c}\Bigg\}^{-1} 
\eea

In view of the one-loop results of $ \chi_t $,  
[Eqs. (\ref{eq:chit_nf2}), (\ref{eq:chit_nf3}), (\ref{eq:chit_nf2p1}), 
and (\ref{eq:chit_nf2p1p1})],       
it would be interesting to see whether the $ \chi_t $ measured 
in lattice QCD with exact chiral symmetry would agree with 
the prediction of ChPT. Most importantly, these one-loop formulas
provide a viable way to determine the  
low-energy constants $ F_\pi $, $ L_6 $, $ L_7 $ and $ L_8 $, 
in addition to the chiral condensate $ \Sigma $ which has already 
been determined \cite{Aoki:2007pw, Chiu:2008jq, Chiu:2008kt} 
using the formula of $ \chi_t $ at the tree level (\ref{eq:chit_tree}).   

Finally, we turn to the second normalized cumulant $ c_4 $. 
At this moment, we only have a formula of $ c_4 $ (\ref{eq:c4_nf_tree})
at the tree level. For $ N_f = 2 $, the ratio $ c_4 / \chi_t = -1/4 $ 
in the isospin limit ($ m_u = m_d $) seems to rule out the 
instanton gas/liquid model which predicts that $ c_4 / \chi_t = -1 $. 
Obviously, it would be interesting to derive a 
formula of $ c_4 $ to the next (non-vanishing) order in ChPT. 

\bigskip
\bigskip

\noindent{\bf Appendix}
\bigskip

In this Appendix, we present a heuristic derivation 
of the relationship between topological susceptibility, 
chiral condensate, and the quark masses, for an arbitrary number of 
(non-degenerate) flavors, in the framework of lattice QCD 
with exact chiral symmetry. Our derivation generalizes that presented 
in Ref. \cite{Chandrasekharan:1998wg} with degenerate flavors. 

Consider the flavor-singlet pseudoscalar $\eta'$
\BAN
\eta'(x) = \frac{1}{N_f} \sum_{i=1}^{N_f} \bar q_i(x) \gamma_5 q_i(x). 
\EAN
Its correlator at zero momentum is
\bea
\label{eq:G_etap}
&& G_{\eta'}(p=0) = \frac{1}{\vol} \sum_{x,y} 
\langle \eta'(x) \eta'^\dagger(y) \rangle \nn 
&=& \frac{1}{\vol N_f^2} \sum_{x,y} \sum_{i,j=1}^{N_f} 
\langle \bar q_i(x) \gamma_{5} q_i(x) \bar q_j(y) \gamma_{5} q_j(y)\rangle \nn
&=& \frac{1}{\vol Z N_f^2} \int [ d U ] \det D(m) e^{-S_g[U]} \times \nn
&& \hspace{10mm} \left\{ 
\sum_{i=1}^{N_f} \Tr[(D_c + m_i)^{-1} \gamma_5 (D_c + m_i)^{-1} \gamma_5]
 -\left( \sum_{i=1}^{N_f} \Tr[(D_c + m_i)^{-1} \gamma_5]\right)^2 \right\} \nn  
&=& \frac{1}{\vol Z N_f^2} \int [ d U ] \det D(m) e^{-S_g[U]}
\left\{ \sum_{i=1}^{N_f} \frac{1}{m_i} \Tr(D_c + m_i)^{-1} -
        \left[\sum_{i=1}^{N_f} \frac{1}{m_i} (n_+ - n_-) \right]^2 \right\}, 
\eea
where $ S_g[U] $ is the gauge action, 
\BAN
\det D(m) &=&  \prod_{i=1}^{N_f} \det[(D_c + m_i)(1+rD_c)^{-1}], \\ 
Z &=& \int [ d U ] \det D(m) e^{-S_g[U]},  
\EAN
and the identity 
\BAN
  \Tr[(D_c + m)^{-1} \gamma_5 (D_c + m)^{-1} \gamma_5]
= \frac{1}{m} \Tr (D_c + m)^{-1}, 
\EAN
has been used in the last equality of (\ref{eq:G_etap}). 
Here $ D_c = D ( 1 - r D)^{-1} $ is the chirally symmetric Dirac operator 
of a Ginsparg-Wilson Dirac operator $ D $ satisfying 
$ D \gamma_5 + \gamma_5 D = 2 r D \gamma_5 D $.

Now taking the thermodynamic limit ($ \vol \to \infty $), and then the 
chiral limit ($ m_{i} \to 0 $), (\ref{eq:G_etap}) gives
\bea
\label{eq:G_etap_p=0}
G_{\eta'}(0) = \frac{1}{N_f^2} 
\left(\sum_{i=1}^{N_f} \frac{1}{m_i} \right) 
\left\{\Sigma-\left(\sum_{i=1}^{N_f} \frac{1}{m_i} \right) \chi_t \right\}, 
\eea
where
\BAN
\Sigma &=& \lim_{m_i \to 0} \lim_{\vol\to\infty} 
              \frac{1}{\vol} \langle \Tr(D_c+m_i)^{-1} \rangle,    \\
\chi_{t} &=& \lim_{m_i \to 0} \lim_{\vol\to\infty} 
             \frac{1}{\vol} \langle (n_{+} - n_{-})^2 \rangle. 
\EAN
If $ \eta' $ stays massive,
then its propagator $ G_{\eta'} \propto m_{\eta'}^{-2}$ must be
non-singular. This implies that the coefficient of the singular factor 
$ \left( \sum_{i=1}^{N_f} \frac{1}{m_i} \right) $ in (\ref{eq:G_etap_p=0}) 
behaves like $ {\cal O}(m) $, i.e.,   
\BAN
\chi_t = \Sigma \left(\sum_{i=1}^{N_f} \frac{1}{m_i} \right)^{-1}, 
\EAN
which agrees with the Leutwyler-Smilga relation.

For $ N_f = 2+1 $ with fixed $ m_s $, (\ref{eq:G_etap_p=0}) is 
modified to 
\bea
\label{eq:G_etap_nf2p1}
G_{\eta'}(0) = \frac{1}{N_f^2} 
\left(\frac{2}{m_{u,d}} \right) 
\left\{ \Sigma \left(1+\frac{m_{u,d}}{2 m_s} 
                       \frac{\langle \bar s s \rangle}{\Sigma} \right)
 - \frac{2}{m_{u,d}} 
   \left(1+\frac{m_{u,d}}{2 m_s} \right)^2 \chi_t \right\}, 
\eea
where
\BAN
\langle \bar s s \rangle
\equiv \lim_{\vol\to\infty} 
       \frac{1}{\vol} \langle \Tr(D_c+m_s)^{-1} \rangle.    
\EAN
In the limit $ m_{u,d} \to 0 $, the coefficient of the singular factor 
$ 2/m_{u,d} $ in (\ref{eq:G_etap_nf2p1}) behaves like 
$ {\cal O}(m_{u,d}) $, i.e.,   
\bea
\label{eq:chit_nf2p1_qcd}
\chi_t 
       &=& \left(   \frac{\Sigma}{m_u} + \frac{\Sigma}{m_d} 
                  + \frac{\langle \bar s s \rangle}{m_s} \right) 
           \left( \frac{1}{m_u} + \frac{1}{m_d} + \frac{1}{m_s} \right)^{-2}, 
\eea
which provides a more physical relationship (between $ \chi_t $, 
$ \Sigma $, $ \langle \bar s s \rangle $, and the quark masses) 
than the Leutwyler-Smilga relation (\ref{eq:LS_nf2p1}), 
since it reduces to (\ref{eq:LS_nf2p1}) 
only in the (unphysical) limit $ \langle \bar s s \rangle = \Sigma $.
 
Now it is straightforward to generalize (\ref{eq:chit_nf2p1_qcd})
to the case $ N_f = 2 + 1 + 1 $,  
\bea
\label{eq:chit_nf2p1p1_qcd}
\chi_t = \left(   \frac{\Sigma}{m_u} + \frac{\Sigma}{m_d}
                + \frac{\langle \bar s s \rangle}{m_s} 
                + \frac{\langle \bar c c \rangle}{m_c} 
         \right) 
         \left( \frac{1}{m_u} + \frac{1}{m_d} + \frac{1}{m_s} + \frac{1}{m_c} 
         \right)^{-2},  
\eea
where
\BAN
\langle \bar c c \rangle
\equiv \lim_{\vol\to\infty} 
       \frac{1}{\vol} \langle \Tr(D_c+m_c)^{-1} \rangle.    
\EAN
It would be interesting to determine $ \Sigma $, 
$ \langle \bar s s \rangle $, and  
$ \langle \bar c c \rangle $ 
with the data of $ \chi_t $ in lattice QCD with exact chiral symmetry.

\begin{acknowledgments}
  We thank Sinya Aoki and Hidenori Fukaya for helpful communications.  
  This work is supported in part by the National Science Council 
  No.~NSC96-2112-M-002-020-MY3, NTU-CQSE (Nos.~97R0066-65,~97R0066-69), 
  and NCTS (No.~NSC98-2119-M-002-001). 
\end{acknowledgments}

\end{document}